\documentclass[12pt]{iopart}
\usepackage{amssymb,amsthm}

\begin{document}
\title{Tensor calculus on noncommutative spaces}
\author{D V Vassilevich}
\address{CMCC, Universidade Federal do ABC, Santo Andr\'e, SP, Brazil}
\address{Department of Theoretical Physics, St Petersburg State University,
St Petersburg, Russia}
\ead{dvassil@gmail.com}
\begin{abstract}It is well known that for a given Poisson structure one has infinitely many
star products related through the Kontsevich gauge transformations. These gauge transformations
have an infinite functional dimension (i.e., correspond to an infinite number of degrees
of freedom per point of the base manifold). We show that on a symplectic manifold this freedom
may be almost completely eliminated if one extends the star product to all tensor fields in
a covariant way and impose some natural conditions on the tensor algebra. The remaining
ambiguity corresponds either to constant renormalizations to the symplectic structure, or
to maps between classically equivalent field theory actions. We also discuss how one can
introduce the Riemannian metric in this approach and the consequences of our results
for noncommutative gravity theories.
\end{abstract}

\section{Introduction}\label{sec-in}
The deformation quantization program was formulated in its modern form in the
papers \cite{BFFLS1,BFFLS2}, see \cite{DS} for an overview. The main part of
this program is the construction of a star product, which is an associative
deformation of the usual point wise product in the direction of a given Poisson
structure. On symplectic manifolds (non-degenerate Poisson structure) the problem
was solved by Fedosov \cite{Fedosov} in a covariant way.
The existence of a star product for an arbitrary Poisson structure was demonstrated by
Kontsevich \cite{Kontsevich}, who also gave a closed formula for this product. To calculate higher
orders of the star it is more convenient to use other methods \cite{Kupriyanov:2008dn}.
The Kontsevich formula is written in a fixed coordinate frame, and, therefore, is not
covariant. A globalization of the Kontsevich product was done by Cattaneo and Felder
\cite{CF}, see also \cite{Dolgushev} for a covariant version of the Kontsevich formality theorem.
More recently, a manifestly covariant universal star product was constructed
\cite{ACG}. Further generalizations of the star product consist in its' extension to
the exterior algebra of differential forms \cite{HoM,BeMa,McCZ}, Lie algebra valued differential
forms \cite{Liev}, and tensor valued differential forms \cite{tenv}. Gauge theories with
covariant star products were considered in \cite{Chaichian:2009kn}.

To a given Poisson structure one can associate infinitely many star products related through
the Kontsevich gauge transformations \cite{Kontsevich}. These gauge transformations depend on
an arbitrary order (formal) differential operator. This means, that the star products are
parametrized by an infinite number of fields, which are coefficients in front of powers of
the derivatives in this operator. In other words, the space of star products has an infinite
functional dimension, corresponding to an infinite number of degrees of freedom per point
of the base manifold.
Although the star products related through the Kontsevich
transformations are equivalent in the sense of deformation quantization, field theories based
on such products are by no means equivalent. The reason is very simple: the Kontsevich gauge
transformations are not gauge symmetries in the field theory sense. To make them local symmetries
of an action one has to add corresponding gauge fields\cite{Pinzul:2007bk}. The number of
such gauge fields is, in general, infinite.
Some part of the gauge freedom
may be used to extract physical fields \cite{Pinzul:2007bk}, but the total ambiguity is enormous
and must be reduced by imposing some natural restrictions on admissible star products.

The problem of symmetries of a physical action on a noncommutative space is one of the central
ones for noncommutative field theories \cite{Szabo}. Due to the presence of the Poisson tensor,
many symmetries are broken, but may be restored by "twisting", which amounts to replacing the
Lie algebra of symmetries by the corresponding Hopf algebra with a twisted coproduct. The idea
of twisting was discussed already in \cite{earlytw}. The first physical symmetry to be realized
in this way was the Poincare one \cite{tP}, which was then followed by other symmetries, as, e.g.,
the diffemorphism \cite{tdiff} and gauge transformations \cite{tgauge}. In this way one can define
practically any symmetry on noncommutative plane. There is, however, a drawback: twisted local
symmetries are not bona fide symmetries. One cannot use them for a gauge-fixing in order to remove
non-physical degrees of freedom or to select a representative of a family of gauge-equivalent
configurations. It is desirable, therefore, to have the symmetries realized in the standard
non-twisted way.

It is natural to address both problems, namely the problem of symmetries and the problem of the
abundance of free parameters, simultaneously. It is well known that the existence of a derivation
satisfying the Leibniz rule is very restrictive. But, the derivative $\nabla_\mu$ maps scalars
to vectors, vectors to rank two tensors, etc. To define the derivative in a consistent way one
has to extend the star product to all tensor fields, thus adding more arbitrariness. Nevertheless,
one gets a possibility to formulate some natural conditions which the algebra of tensors should satisfy.
These conditions appear to be restrictive enough to remove practically all ambiguities in the
star product.

This paper is organized as follows. In the next section we re-introduce a covariant star product
\cite{BFFLS1} and remind some basic properties of its extension to the tensor fields \cite{Vassilevich:2009cb}.
Then, in Sec.\ \ref{sec-gf}, we remind the structure of the Kontsevich gauge transformations, impose
some conditions on the star product, and show that the ambiguity is thus reduced to a space of
a finite functional dimension. Next (Sec. \ref{sec-twist}), we show that all products satisfying
the conditions of Sec.\
\ref{sec-gf}, can be realized through a twist. In Sec.\ \ref{sec-clo} we demonstrate that
for closed star products the gauge freedom is reduced further, and that practically all remaining
gauge transformations are maps between classically equivalent actions. What remains, is just
constant renormalization of the symplectic structure (i.e., the freedom to add covariantly constant
terms to the symplectic structure with higher orders of the deformation parameter) and a choice
of a flat torsion-free symplectic connection. In Sec.\ \ref{sec-met} we discuss some implications
for noncommutative gravity theories. Finally, Sec.\ \ref{sec-con} contains concluding remarks.
In this section we also argue that the set of the requirements that we impose on the star product
is a natural replacement of locality in the noncommutative setting. We support these arguments by
considering the case of vanishing Poisson structure in Appendix \ref{appA}.

\section{The star product}\label{sec-sta}
Consider a symplectic manifold $M$ with a symplectic form
$\omega_{\mu\nu}$ (with the Poisson bivector $\omega^{\mu\nu}$
being its' inverse, $\omega_{\mu\nu}\omega^{\nu\rho}=\delta_\mu^\rho$).
Let $TM$ be a tangent bundle, and $T^*M$ be a
cotangent bundle. Let $\alpha_{n,m}$ be a tensor field,
$\alpha_{n,m}\in TM^n \otimes T^*M^m\equiv T^{n,m}$. This means,
$\alpha_{n,m}$ has $n$ contravariant and $m$ covariant indices.

Let us choose a Christoffel symbol
on $M$ such that the symplectic form is covariantly constant,
\begin{equation}
\nabla_{\mu}\omega_{\nu\rho}=
\partial_\mu \omega_{\nu\rho}-\Gamma_{\mu\nu}^{\sigma}\omega_{\sigma\rho}
-\Gamma_{\mu\rho}^\sigma \omega_{\nu\sigma}=0.\label{cco}
\end{equation}
Therefore, ${M}$ becomes a Fedosov manifold \cite{GRS}. Let us suppose
that this connection is flat and torsion-free, i.e.,
\begin{equation}
 [\nabla_\mu,\nabla_\nu]=0.\label{flatcon}
\end{equation}
Locally, one can choose a coordinate system such that $\omega^{\mu\nu}=const.$ and $\Gamma_{\mu\nu}^\sigma=0$.
We shall call such coordinates the Darboux coordinates.

One can then consider the algebra of formal power series $T[[h]]$, where
$T\equiv \bigoplus_{n,m} T^{n,m}$, with $h$ being a deformation parameter, and
define a covariant star product \cite{BFFLS1} (it was recently used in \cite{Vassilevich:2009cb}
in the context of NC gravity)
\begin{equation}
\alpha\star\beta =\sum_k \frac{h^k}{k!} \omega^{\mu_1\nu_1}
\dots \omega^{\mu_k\nu_k} (\nabla_{\mu_1} \dots \nabla_{\mu_k}
\alpha) \cdot (\nabla_{\nu_1}\dots \nabla_{\nu_k} \beta)\,.
\label{dvstar}
\end{equation}
We stress that this product respects the diffemorphism symmetry, i.e., it indeed
maps tensors to tensors. Apart of this, it has the following obvious properties
\begin{itemize}
\item[S1.]
\begin{equation*}
\alpha \star \beta = \alpha \beta +\sum_{k=1}^\infty h^k
C_k(\alpha,\beta),
\end{equation*}
where $C_k$ are bilinear differential operators.
\item[S2.]
Associativity:
\begin{equation*}
\alpha \star (\beta\star\gamma)=(\alpha\star\beta)\star\gamma\,.
\end{equation*}
\item[S3.] The order $h$ term is a Poisson bracket,
$C_1(\alpha,\beta)=\{\alpha,\beta\}$.
\item[S4.] Stability on covariantly constant tensors: $\alpha\star\beta=\alpha \cdot \beta$
if $\nabla\alpha=0$ or $\nabla\beta=0$.
\item[S5.] The Moyal symmetry:
\begin{equation*}
C_k(\alpha,\beta)=(-1)^k C_k(\beta,\alpha).\end{equation*}
\item[S6.] Derivation:
\begin{equation*}
\nabla \alpha\star\beta = (\nabla\alpha)\star\beta + \alpha\star
(\nabla\beta).\end{equation*}
\end{itemize}

The list above is an extension of the requirements imposed on the star products
of forms \cite{McCZ}, except for S6, which was not requested in \cite{McCZ}.
Instead of S4 one usually considers a weaker property, namely, that the unit function
is the unity of the algebra, $1\star \alpha = \alpha \star 1=\alpha$. From the physical
point of view, S4 means that for slowly varying fields the star product should look
as the usual point wise product.
The property S3 \textit{defines} a Poisson bracket on $T$
\begin{equation}
\{\alpha,\beta\}=\omega^{\mu\nu} \nabla_\mu\alpha \cdot \nabla_\nu\beta \label{pbr}
\end{equation}
having the following properties.
\begin{itemize}
\item[P1.] Bracket degree:
\begin{equation*}
\{ \alpha_{n_1,m_1},\beta_{n_2,m_2}\} \in T^{n_1+n_2,m_1+m_2}.
\end{equation*}
\item[P2.] Antisymmetry
\begin{equation*}
\{ \alpha,\beta\}=-\{ \beta,\alpha\}.
\end{equation*}
\item[P3.] Product rule:
\begin{equation*}
\{ \alpha,\beta\gamma\}=\{\alpha,\beta\} \gamma + \beta
\{\alpha,\gamma\}.
\end{equation*}
\item[P4.] There is a covariant derivative $\nabla: T^{n,m}\to
T^{n,m+1}$ such that $[\nabla_\mu,\nabla_\nu]=0$ and
\begin{equation*}
\nabla \{ \alpha ,\beta \}=\{ \nabla\alpha,\beta\} +
\{\alpha,\nabla\beta\}.\end{equation*}
\item[P5.] Jacobi identity:
\begin{equation*}
\{ \alpha,\{ \beta,\gamma\}\} + \{\gamma,\{ \alpha,\beta\}\} +
\{\beta,\{ \gamma,\alpha\}\}=0.\end{equation*}

\end{itemize}

The properties P1-P5 can be demonstrated by expanding corresponding conditions S1-S6
and picking up appropriate power of $h$, or by using the explicit formula (\ref{pbr}).
Again, P1-P6 are extensions to arbitrary tensors of the requirements on the Poisson
structure of differential forms \cite{Chu:1997ik,McCZ}.

Let us make an important remark. It is known that the Leibniz rule S6 is very
restrictive. However, since $\nabla$ maps a scalar to e vector, imposing S6 in
a way compatible with covariance requires first to extend the star product to
tensor fields, which enlarges the freedom in the choice of the star product.

In physical application the deformation parameter $h$ is imaginary, $\bar h=-h$.
Then S5 yields that the star product is Hermitian,
\begin{equation}
\overline{(\alpha\star \beta)}=\bar \beta \star \bar \alpha\,,\label{her}
\end{equation}
Since the symplectic structure is covariantly constant,
\begin{equation}
\omega^{\mu\nu}\star \alpha =\omega^{\mu\nu}\cdot \alpha \label{ost}\,,
\end{equation}
i.e., $\omega^{\mu\nu}$ is central in the corresponding commutator
algebra.

There is a natural integration measure (see \cite{Felder:2000nc}
for a detailed discussion of traces in relation to star products)
\begin{equation}
d\mu (x)=(\det (\omega^{\mu\nu}))^{-\frac 12}\, dx\,,
\label{intme}
\end{equation}
with respect to which the star product
of tensors is closed provided all indices are contracted in pairs,
\begin{equation}
\int_{{M}} d\mu (x) \alpha_{\mu\nu\dots\rho}\star \beta^{\mu\nu\dots\rho}=
\int_{{M}} d\mu (x) \alpha_{\mu\nu\dots\rho}\cdot \beta^{\mu\nu\dots\rho}\,.
\label{clo}
\end{equation}
In other words, the property (\ref{clo}) is valid when the integrand
is diffeomorphism invariant.

\section{Gauge freedom}\label{sec-gf}
\subsection{Gauge transformations}\label{sec-gtr}
A star product corresponding to a given Poisson structure is not unique. The arbitrariness
in the choice of a star product is described by the Kontsevich \cite{Kontsevich}
gauge transformation
$\star\to\star'$:
\begin{equation}
\alpha\star'\beta = D^{-1} (D\alpha \star D\beta),\label{gatr}
\end{equation}
where
\begin{equation}
D=1+hL \,.\label{Dtau}
\end{equation}
Here $L$ is a formal differential operator, i.e, it is
sum of differential operators of arbitrary order
\begin{equation}
L = \sum_{k=1}^{\infty} L^{\mu_1 \dots \mu_k}\nabla_{\mu_1}\dots\nabla_{\mu_k}\,,
\label{tausum}
\end{equation}
and each $L^{\mu_1 \dots \mu_k}$ contains non-negative powers of the deformation parameter $h$.
Eq.\ (\ref{tausum}) is nothing else than a covariantization of corresponding formula in \cite{Kontsevich}. Here we also like to mention an interpretation of the gauge freedom
in the framework of the Fedosov approach given in \cite{AK}.

We like to exclude from (\ref{gatr}) the transformations which leave the star product
invariant. They correspond to inner automorphisms of the algebra
\begin{equation}
\alpha \to \alpha_\varphi =
e^{\varphi}_\star \star \alpha \star e_\star^{-\varphi}\label{inaut}
\end{equation}
where $e^{\varphi}_\star =1 +\varphi +
(1/2) \varphi\star\varphi + \dots$ is the star-exponent. $\varphi$ is a scalar which can be expanded
in non-negative powers of the deformation parameter $h$. In physical units $h$ is imaginary, and
$\varphi$ is imaginary as well, so that we have a $U(1)_\star$ gauge group (see \cite{Szabo}).
It is easy to check that the transformations (\ref{inaut}) indeed do not change the star product,
$\alpha_\varphi \star \beta_\varphi = (\alpha\star\beta)_\varphi$, and that they can be represented in the
form (\ref{gatr}) with
\begin{equation}
hL (\varphi)=2h\omega^{\mu\nu} (\nabla_\mu \varphi) \nabla_\nu + \dots \,,\label{tauvarp}
\end{equation}
where we omitted higher derivative terms. Therefore, in what follows we exclude from the gauge
transformations the terms whose linear part has the form (\ref{tauvarp}).

\subsection{Reduction of the gauge freedom}\label{sec-red}
Let us now study the following question. Suppose we have a Fedosov manifold with a symplectic structure
$\omega^{\mu\nu}$ and a flat torsionless symplectic connection $\nabla$.  What are the star products
which preserve the rank of the tensors, $T^{n_1,m_1}\star T^{n_2,m_2}\subset T^{n_1+n_2,m_1+m_2}$,
and satisfy the requirements S1-S6 with the Poisson bracket defined in Eq.\ (\ref{pbr})? Since the Poisson
structure is given by (\ref{pbr}), such products belong to the family consisting of
the star product (\ref{dvstar}) and the products gauge-equivalent to (\ref{dvstar}).

Interestingly, ``physical uniqueness'' of the star product (\ref{dvstar}) for scalars
was demonstrated already in
\cite{BFFLS1}. It was shown, that if one has only the symplectic structure and the flat covariant
derivative, the star product must be a sum of the same terms as
on the right hand side of (\ref{dvstar})
but with possibly different coefficients. Correct coefficients are then fixed by requiring associativity
of the product.

Here we shall allow for arbitrary structures to appear and then reduce the gauge freedom by using S1-S6.
This will be done in three step. First, we shall show that $L$ does not depend on the tensor degree.
Second, we shall exclude the first order terms in $L$ modulo "constant renormalizations" of
the symplectic structure (defined below). Third, we will show that the higher order terms in $L$
(i.e., with two and more derivatives) have to be covariantly constant. Each step will include two
sub-steps. We shall begin with analyzing the linearized gauge transformations $\star\to\star'_1$,
where
\begin{equation}
\alpha\star^{'}_1\beta=\alpha\star\beta -hL(\alpha\star\beta) +h(L\alpha \star\beta)
+h(\alpha\star L\beta) \,.\label{lins}
\end{equation}
Afterwards, the results will be extended to full non-linear gauge transformations (\ref{gatr}).
Perhaps, this is not the shortest way to prove the main result of this section, but it is
probably a more pedagogical one.

A priori, each $L^{\mu_1\dots\mu_k}$ is a local linear map on the space of the tensors. According to our
assumption, $L$ must preserve the rank of the tensors, i.e., it maps $T^{n,m}$ to $T^{n,m}$. Let us denote
the restriction of $L^{\mu_1\dots\mu_k}$ to $T^{m,n}$ as $L^{\mu_1\dots\mu_k}_{(n,m)}$, and let us check
under which conditions the property S4 holds. Take $\alpha_{n,m}$ being covariantly constant, $\nabla\alpha_{n,m}=0$. Then
\begin{eqnarray}
&&0=\alpha_{n,m}\star'_1\beta_{0,0}-\alpha_{n,m}\cdot\beta_{0,0}\label{check4}\\
&&\ \ =h\sum_{k} \left( -L^{\mu_1\dots\mu_k}_{(n,m)} \alpha_{n,m}
\nabla_{\mu_1}\dots \nabla_{\mu_k} \beta_{0,0} +
\alpha_{(n,m)}L_{(0,0)}^{\mu_1\dots\mu_k}\nabla_{\mu_1}\dots \nabla_{\mu_k} \beta_{0,0} \right)
\nonumber
\end{eqnarray}
as a consequence of  S4 in the infinitesimal setting, Eq.\ (\ref{lins}).
This last equation is local and valid for an arbitrary scalar
$\beta_{0,0}$. Although $\alpha_{(n,m)}$ above is restricted to being covariantly constant, it can
have arbitrary value at a given point. Therefore, (\ref{check4}) gives enough conditions to conclude that
\begin{equation}
L^{\mu_1\dots\mu_k}_{(n,m)}=
{\mathrm{I}}_{(n,m)}\, L^{\mu_1\dots\mu_k}_{(0,0)} \,,\label{res4}
\end{equation}
where ${\mathrm{I}}_{(n,m)}$ is the unit operator on $T^{(n,m)}$. To extend this result
to full gauge transformations, let us fix a pair $(n,m)$ and consider the {\textit{lowest}}
number $k$ for which (\ref{res4}) {\textit{does not}} hold. Then, for a covariantly constant
$\alpha_{n,m}$, the terms with $k$ derivatives on $\beta_{0,0}$ in
$ \alpha_{n,m}\star'\beta_{0,0}-\alpha_{n,m}\cdot\beta_{0,0}$ (not restricting this difference
to the linear order any more) look precisely as the expression in the brackets on the second
line of Eq.\ (\ref{check4}). Due to the arbitrariness of $\beta_{0,0}$ this leads to a
contradiction. We conclude, that (\ref{res4}) holds also if $\star'$ is related to $\star$
by a full gauge transformation (\ref{gatr}).

Next, let us study the restrictions on $L$ following from the Leibniz rule S6.
Locally, we can choose a Darboux coordinate system, so that $\omega^{\mu\nu}$
is constant, and $\nabla=\partial$. We start with the first order part $L=L^\mu\partial_\mu$.
Again, we start with the case when $\star'$ and $\star$ are related by a linearized gauge
transformation (\ref{lins}).
The local ($h^0$) part of the product (\ref{dvstar}) is invariant under the gauge transformation
with the first-order $L$, which is a consequence of Eq.\ (\ref{res4}). The terms with lowest
order derivatives in the transformed star product are coming from the $h^1$ part of (\ref{dvstar}):
\begin{eqnarray}
&&\alpha\star^{'}_1\beta - \alpha\star\beta\nonumber\\
&&\quad = h^2[-L^\mu\partial_\mu(\partial_\rho \alpha \partial_\sigma \beta \omega^{\rho\sigma})
+(\partial_\rho L^\mu \partial_\mu \alpha)(\partial_\rho\beta)\omega^{\rho\sigma}
 +(\partial_\rho  \alpha)(\partial_\rho L^\mu\partial_\mu \beta)\omega^{\rho\sigma}]\nonumber\\
&&\quad =h[\partial_\rho\alpha\, \partial_\sigma \beta\, \delta\omega^{\rho\sigma}]\,,
\label{L1tra}
\end{eqnarray}
where
\begin{equation}
\delta\omega^{\rho\sigma}=h[(\partial_\mu L^\rho)\omega^{\mu\sigma}+
\omega^{\rho\mu}(\partial_\mu L^\sigma)]\label{dom1}
\end{equation}
is nothing else than the variation of $\omega^{\rho\sigma}$ under infinitesimal diffeomorphisms
generated by the vector field $L^{\mu}$. The transformation which do not change $\omega^{\rho\sigma}$
(symplectomorphisms)
have already been excluded, see sec.\ \ref{sec-gtr}, so that for the remaining $L^\mu$ the right
hand side of (\ref{dom1}) is non-zero. S6 immediately yields
\begin{equation}
\partial_\nu\, \delta\omega^{\rho\sigma}=0\,.\label{dsicon}
\end{equation}
This means that the allowed gauge transformation are the ones which add to the Poisson structure
arbitrary constant terms containing at least one power of the deformation parameter
$h$. It is natural to call these gauge transformations constant
renormalizations of the Poisson structure.

Let us extend this result beyond the linearized gauge transformations (\ref{lins}).  To this end
we write in the Darboux coordinates the product $\star'$ in a form similar to S1
\begin{equation}
\alpha\star' \beta = \alpha\cdot\beta +\sum_{j,l} B_{(j),(l)} (\partial^{(j)}\alpha )\cdot
(\partial^{(l)}\beta),\label{staB}
\end{equation}
where $(j)$ and $(l)$ are multi-indices, $\partial^{(j)}\equiv
\partial_{\mu_1}\dots \partial_{\mu_j}$. Due to (\ref{res4}) the coefficient functions
$B_{(j),(l)}$ do not depend on the tensor degree of $\alpha$ and $\beta$. Moreover,
because of S4, $B_{(0),(l)}=B_{(j),(0)}=0$, and the sum in (\ref{staB}) starts with
$j=l=1$. Next we write
\begin{equation}
\partial_\nu (\alpha \star' \beta) - (\partial_\nu \alpha \star' \beta)
-(\alpha\star'\partial_\nu\beta)
=\sum_{j,l} (\partial_\nu B_{(j),(l)}) (\partial^{(j)}\alpha )\cdot
(\partial^{(l)}\beta)
 \label{Lei}
\end{equation}
for arbitrary $\alpha$ and $\beta$. We conclude, that to satisfy S6, the coefficient functions
$B_{(j),(l)}$ must be constant in the Darboux coordinates (or covariantly constant in an arbitrary
coordinate system).

For the future use it is convenient to introduce a ``renormalized'' star product
\begin{equation}
\alpha\star_R\beta =\sum_k \frac{h^k}{k!} \omega^{\mu_1\nu_1}_R
\dots \omega^{\mu_k\nu_k}_R (\nabla_{\mu_1} \dots \nabla_{\mu_k}
\alpha) \cdot (\nabla_{\nu_1}\dots \nabla_{\nu_k} \beta)\,,
\label{Rstar}
\end{equation}
which depends on a ``renormalized'' symplectic structure
\begin{equation}
\omega_R^{\mu\nu}=\omega^{\mu\nu}+h^2 \omega^{\mu\nu}_1+h^4\omega^{\mu\nu}_2+\dots
\label{Rom}
\end{equation}
It is supposed, that all correction terms $\omega^{\mu\nu}_j$ are covariantly constant,
\begin{equation}
 \nabla_\rho \omega^{\mu\nu}_j=0. \label{ccRom}
\end{equation}
Note, that odd powers of the deformation parameter $h$ in (\ref{Rom}) are excluded by the Moyal
symmetry requirement S5. Since the $h^0$ part in (\ref{Rom}) is unchanged, the product $\star_R$
is equivalent to $\star$ in the sense of deformation quantization. Other equivalent star products
$\star'$ are obtained from $\star_R$ by means of the gauge transformations (\ref{gatr}). However,
one has to exclude the transformations which do not change the antisymmetric part of $B_{(1),(1)}$.
This has to be done for the following reasons. First, one should avoid double counting of the
degrees of freedom, which are already included in $\omega_j^{\mu\nu}$, i.e. in the \emph{constant}
antisymmetric part of $B_{(1),(1)}$. Second, a \emph{non-constant} $B_{(1),(1)}$ contradicts S4,
as we have just seen above. One can easily see, that this excludes the gauge transformations having
a non-zero first-order part $L^\mu$. To demonstrate this, let us expand $L^\mu$ in a series of
$h$ and pick up the term with the lowest power of $h$. Then, to this order of $h$, the \emph{full}
gauge transformation (including also arbitrary higher derivative terms) of the antisymmetric part
of $B_{(1),(1)}$ is given by the \emph{linearized} expression, c.f. right hand side of (\ref{L1tra}).
The linearized transformations have already been analyzed above,
yielding that this lowest order part of $L^\mu$ must vanish. Consequently, $L^\mu=0$ to all orders.

We have reduced the set of admissible star products to the products, which are obtained from
$\star_R$ by means of the gauge transformations with vanishing first order part $L^\mu$.
Moreover, we know that other $L^{\mu_1\dots\mu_k}$ are proportional to the unit operator on
each $T^{n,m}$ and do not depend on the tensor degree. The coefficient function $B_{(j),(l)}$
have been shown to be constants, which does not yet imply that $L^{\mu_1\dots\mu_k}$ must be
constants as well.

Next, let us analyze the higher derivative terms in $L$ restricting ourselves for a while
to the linear order in $L$.
Such terms change already the local, $h^0$, part of the star product.
For $L=L^{\mu_1\dots\mu_n}\nabla_{\mu_1}\dots\nabla_{\mu_n}$ we have in the Darboux coordinates
\begin{equation}
\alpha\star'_1\beta - \alpha\star\beta = -h\sum_{k=1}^{n-1}C_n^k L^{\mu_1\dots \mu_n}
(\partial_{\mu_1}\dots \partial_{\mu_k}\alpha)(\partial_{\mu_{k+1}}\dots\partial_{\mu_n}
\beta) \label{traLn}
\end{equation}
where $C_n^k$ are binomial coefficients. The terms with a larger number of derivatives
acting on $\alpha$ and $\beta$ have been omitted. Substitution of (\ref{traLn}) in the
Leibniz rule S6 yields
\begin{eqnarray}
&&\partial_\nu (\alpha \star'_1 \beta) - (\partial_\nu \alpha \star'_1 \beta)
-(\alpha\star'_1\partial_\nu\beta)\nonumber\\
&&\quad =-h(\partial_\nu L^{\mu_1\dots \mu_n}) \sum_{k=1}^{n-1}C_n^k
(\partial_{\mu_1}\dots \partial_{\mu_k}\alpha)(\partial_{\mu_{k+1}}\dots\partial_{\mu_n}
\beta) \label{LnS6}
\end{eqnarray}
up to higher derivative terms.

Let us pick up the smallest $n$ such that $L^{\mu_1\dots \mu_n}$
is not (covariantly) constant. Then, the terms with the lowest number of derivatives in
$\nabla (\alpha \star'_1 \beta) - (\nabla \alpha \star'_1 \beta)
-(\alpha\star'_1\nabla\beta)$ are given in the Darboux coordinates precisely by the right
hand side of (\ref{LnS6}). Since $\alpha$ and $\beta$ are arbitrary, we conclude that
\begin{equation}
\nabla_\nu L^{\mu_1\dots \mu_n}=0\,.\label{Lncc}
\end{equation}
Next, we exclude non-constant part of $L^{\mu_1\dots \mu_n\mu_{n+1}}$, and so on.

Note, that no cancellation can occur between the terms (\ref{L1tra}) and (\ref{traLn})
since the former are antisymmetric in $\alpha\leftrightarrow\beta$, while the latter
are symmetric. According to the Moyal symmetry requirement, these two types of the
terms can only appear accompanied by different powers of the deformation parameter
$h$.

Extension of (\ref{Lncc}) to full gauge transformations is straightforward. It is enough
to pick up for each $n$ the lowest power of the deformation parameter $h$ at which
(\ref{Lncc}) is not satisfied, and obtain a contradiction to S4.

There is another, rather obvious, restriction on the coefficients $L^{\mu_1\dots\mu_n}$
which follows from the Moyal symmetry requirement S5. Namely, these coefficients are
allowed to contain odd powers of the deformation parameter $h$ only.

We have just demonstrated the following statement.

{\bf Theorem \ref{sec-gf}.1.} \emph{
Let $M$ be a symplectic manifold with a symplectic structure $\omega^{\mu\nu}$ and a flat torsionless
connection $\nabla$. Any covariant star product on the space of tensor fields over $M$ satisfying
S1-S6 with the Poisson bracket (\ref{pbr}) can be represented as
\begin{equation}
\alpha\star_N\beta = D^{-1}(D\alpha\star_RD\beta)\,,\label{starN}
\end{equation}
where $\alpha,\beta \in T$, the product $\star_R$ is defined in (\ref{Rstar}), and the coefficients
$L^{\mu_1\dots\mu_n}$ in the gauge
operator $D$ are covariantly constant and contain odd powers of the deformations parameter $h$. The coefficient
with n=1 vanishes.}

We shall call the star product (\ref{starN}) the natural star product.

The freedom remaining in the definition of $\star_N$ is that of choosing a single scalar
field\footnote{This can be seen by identifying constant coefficients $L^{\mu_1\dots\mu_n}$
with coefficients in the Taylor expansion of a scalar field.}
depending also on $h$. Therefore, the space of natural star products for a given $\omega^{\mu\nu}$
and $\nabla$ has a functional dimension one. Let us remind, that the space of arbitrary
products $\star'$ corresponding to given $\omega^{\mu\nu}$
and $\nabla$, eq.\ (\ref{gatr}), has an \emph{infinite} functional dimension.

In Appendix \ref{appA} we explain how the arguments of this Section work in the
commutative case $\omega^{\mu\nu}=0$. 

\section{Twist representation of the star product}\label{sec-twist}
In this section, we are going to demonstrate that any natural star product (\ref{starN})
can be represented
through a twist on a suitable Hopf algebra. A rather complete survey on the
Hopf algebras can be found in the monographs \cite{Montgomery} and \cite{Majid}. A minimal set of
necessary information is contained in \cite{Szabo}.

Consider a Lie algebra $A$. One can make out of the universal enveloping
algebra $\mathcal{U}(A)$ a Hopf algebra $H$ by introducing a primitive
coproduct $\Delta_0:\ H\to H\otimes H$ such that $\Delta_0(1)=1\otimes 1$
and $\Delta_0(a)=1\otimes a + a \otimes 1$, where $a$ is a generator of $A$.
A counit $\varepsilon: H\to \mathbb{C}$ is defined by the relations
$\varepsilon (a)=0$, $\varepsilon (1)=1$, and an antipode $S:H\to H$
satisfies $S(1)=1$, $S(a)=-a$.

Consider an invertible\footnote{One can consider also non-invertible twist elements,
see \cite{Kur}, though this leads to some technical complications.}
element $\mathcal{F}\in H\otimes H$. If it satisfies
the conditions
\begin{eqnarray}
&& (\mathcal{F}\otimes 1) (\Delta_0 \otimes 1) \mathcal{F}=
(1\otimes \mathcal{F} )(1\otimes \Delta_0)\mathcal{F}\,,\label{twist1}\\
&&(\varepsilon\otimes 1) \mathcal{F}=1=(1\otimes \varepsilon) \mathcal{F}\,,
\label{twist2}
\end{eqnarray}
it is called a twist \cite{Resh}, and it defines a new twisted Hopf algebra.

If the second equation (\ref{twist2}) only is satisfied by $\mathcal{F}$, the twisted
algebra is quasi-Hopf, and not Hopf (see, e.g., \cite{Majid}, Theorem 2.4.2),
meaning that the co-associativity is lost. Although quasi-Hopf algebras were recently
considered in the context of noncommutative quantum field theory \cite{Bal}, we do not
consider such a possibility here and assume that $\mathcal{F}$ satisfies both the
equations (\ref{twist1}) and (\ref{twist2}).

Suppose that $H$ acts on $T$. There is a commutative point wise product $\mu_0$
on $T$, $\mu_0(\alpha\otimes\beta)=\alpha\cdot\beta$. By using the twist, one
can define a new associative product $\mu_{\mathcal{F}}=
\mu_0 \circ \mathcal{F}^{-1}$, i.e.,  as
\begin{equation}
\mu_{\mathcal{F}}(\alpha\otimes\beta)=
\mu_0(\mathcal{F}^{-1}( \alpha\otimes\beta)).
\label{muF}
\end{equation}
We are going to demonstrate that each natural star product can be represented in this
form
\begin{equation}
\alpha\star_N\beta = \mu_0 (\mathcal{F}_N^{-1} (\alpha\otimes\beta))\label{sNFN}
\end{equation}
for some twist $\mathcal{F}_N$.

Let us pass to the Darboux coordinate system (transition to general coordinates will be
considered at the end of this section). Let $A$ be an abelian algebra generated by the
partial derivatives $\partial_\mu$, and $H$ be the corresponding universal enveloping algebra
with a primitive coproduct $\Delta_0$ and the counit and antipode as defined above. In the
Darboux coordinates the product $\star_R$ coincides with the Moyal product with a renormalized
symplectic structure $\omega_R$. Therefore, it can be represented through the standard
(Moyal) twist
\begin{equation}
\mathcal{F}_R=\exp \left( -h \omega_R^{\mu\nu} \partial_\mu \otimes \partial_\nu \right)
\label{FR}
\end{equation}
To proceed with generic $\star_N$, let us represent the admissible gauge transformations
in an exponential form
\begin{equation}
D=\exp \left( h \sum_{n=2}^\infty l^{\mu_1\dots\mu_n}\partial_{\mu_1}\dots\partial_{\mu_n}
\right) \label{Dex}
\end{equation}
with constant coefficients $l^{\mu_1\dots\mu_n}$. Obviously, $D$ is an element of $H$. One
can find $\mathcal{F}_N\in H\otimes H$ such that (\ref{sNFN}) is satisfied:
\begin{eqnarray}
&&\mathcal{F}_N=\mathcal{F}_R\tilde \mathcal{F},\label{FNFR}\\
&&\tilde \mathcal{F}^{-1}=\Delta_0 (D^{-1}) (D\otimes D) \,.\label{FDDD}
\end{eqnarray}
Note that $\tilde\mathcal{F}$ and $\mathcal{F}_R$ commute. Equation (\ref{FNFR}) follows from the
fact that $\star_N$ is a gauge transformation of $\star_R$. To derive Eq.\ (\ref{FDDD}) one observes
that the primitive coproduct corresponds to the usual Leibniz rule, i.e., $\partial_\mu (\alpha\cdot\beta)
=\mu_0(\Delta_0(\partial_\mu)(\alpha\otimes\beta))$. Moreover, coproduct is an algebra morphism. Consequently,
$\Delta_0$ in (\ref{FDDD}) distributes the derivatives contained in $D^{-1}$ in the tensor product
$D\otimes D$. Explicitly,
\begin{equation}
\tilde\mathcal{F}=\exp\left( h\sum_{n=2}^{\infty} l^{\mu_1\dots\mu_n}
\sum_{k=1}^{n-1} C_n^k \partial_{\mu_1}\dots\partial_{\mu_k}\otimes
\partial_{\mu_{k+1}}\dots\partial_{\mu_n} \right)\label{expFt}
\end{equation}
(cf Eq. (\ref{traLn})).

Next, we have to show that $\mathcal{F}_N$ satisfies both (\ref{twist1}) and (\ref{twist2}).
Equation (\ref{twist2}) is straightforward. Since $\varepsilon (\partial_\mu)=0$, only the
unit elements in $\tilde\mathcal{F}$ and $\mathcal{F}_R$ contribute to $(\varepsilon\otimes 1)\mathcal{F}$
and $(1\otimes \varepsilon)\mathcal{F}$. Eq.\ (\ref{twist2}) follows immediately. By using certain
commutativity properties and the fact that $\mathcal{F}_R$ satisfies Eq.\ (\ref{twist1}), one can
show, that $\mathcal{F}_N$ satisfies (\ref{twist1}) if and only if $\tilde\mathcal{F}$ satisfies
the same equation with the primitive coproduct $\Delta_0$:
\begin{equation}
(\tilde \mathcal{F}\otimes 1) (\Delta_0 \otimes 1)\tilde \mathcal{F}=
(1\otimes \tilde \mathcal{F} )(1\otimes \Delta_0)\tilde\mathcal{F}\,.\label{t1til}
\end{equation}
Eq.\ (\ref{t1til}) can be checked directly. Indeed, one can bring left and right hand sides
of (\ref{t1til}) to the form
\begin{equation*}
\exp\left( h \sum_{n=2}^\infty l^{\mu_1\dots\mu_n} \sum_{i+j+k=n}
\frac{n!}{i!j!k!}\, \partial_{\mu_1}\dots\partial_{\mu_i}\otimes
\partial_{\mu_{i+1}}\dots\partial_{\mu_{i+j}}\otimes
\partial_{\mu_{i+j+1}}\dots\partial_{\mu_n} \right)
\end{equation*}

Note, that the twist representation is not a universal property of the star products.
For generic star product such a representation is not known.

Having defined the twist $\mathcal{F}_N$, we can define a new coproduct, $\Delta=
\mathcal{F}_N \Delta_0 \mathcal{F}_N^{-1}$, and, after suitably extending the algebra $A$,
we can also define twisted symmetries (Poicare, diffeomorphism, Yang-Mills, etc.).
Twisting the diffeomorphism seems unnecessary, since we have these transformation realized
in the standard way. Here we repeat the interpretation of twisted local symmetries suggested
in \cite{Vassilevich:2007jg,Vassilevich:2009cb}: twisted diffeomorphisms is what remains
from the standard diffeomorphism symmetry when $\omega$ and $\nabla$ are gauge fixed to
some given values. It is also possible, that twisted symmetries are effective low-energy
symmetries when the noncommutativity is defined and fixed by some high-energy effects.

In an arbitrary coordinate system the tensors $\omega^{\mu\nu}$ and $l^{\mu_1\dots\mu_n}$
are not constants. Therefore, (\ref{FR}) and (\ref{expFt}) are not linear combinations
of elements of the tensor square of $\mathcal{U}(A)$. To overcome this difficulty, one introduces
a set of vector fields $\xi^\mu_a$, which are analogs of the vielbein fields of Riemannian
geometry, that $\omega^{\mu\nu}=\xi^\mu_a \hat \omega^{ab} \xi_b^\nu$ with a constant
$\hat\omega^{ab}$. These fields locally describe the transition to a Darboux coordinate
system. The rest is obvious. For example, $\omega^{\mu\nu}\partial_\mu\otimes\partial_\nu$
is replaced by $\hat\omega^{ab}(\xi_a^\mu\nabla_\mu)\otimes(\xi_b^\nu\nabla_\nu)$. The Hopf
algebra is then generated by the fields $\xi_a^\mu\nabla_\mu$.
\section{Integration and closeness}\label{sec-clo}
Let us check which of the natural star products are also closed, i.e. when
the Equation (\ref{clo}) is satisfied by $\star=\star_N$. Let $\alpha\in T^{m,n}$
and $\beta\in T^{n,m}$. The following chain of transformations is obvious in
the Darboux coordinates. We suppose that all indices of $\alpha$ and $\beta$ are
contracted in pairs, making the integrals below diffeomorphism invariant.
\begin{eqnarray}
&&\int_M d\mu(x) \alpha \star_N \beta = \int_M d\mu(x) D^{-1} ((D\alpha) \star_R (D \beta))
\nonumber\\
&&=\int_M d\mu(x) (D\alpha) \star_R (D\beta)= \int_M d\mu(x) (D\alpha) \cdot (D\beta)\nonumber\\
&&=\int_M d\mu(x) \alpha \cdot (D^\dag D\beta)\,,\label{1clo}
\end{eqnarray}
where $D$ is given by (\ref{Dex}) with (covariantly) constant coefficient. To obtain the second
line we neglect all total derivative terms coming from $D^{-1}$ and used closeness of $\star_R$,
which is obvious. The operator $D^\dag$ differs from $D$ by the signs in front of $l^{\mu_1\dots\mu_n}$
with odd $n$, i.e.
\begin{equation}
D^\dag D=\exp \left( 2h \sum_{n=1}^\infty l^{\mu_1\dots\mu_{2n}}\partial_{\mu_1}\dots\partial_{\mu_{2n}}
\right)\,. \label{DdD}
\end{equation}
The integral (\ref{1clo}) has to be equal to $\int d\mu(x)\alpha\cdot\beta$ for fairly arbitrary $\alpha$
and $\beta$. This implies that $D$ has to be unitary, $D^\dag D=1$. In other words, $\star_N$ is closed
iff
\begin{equation}
l^{\mu_1\dots\mu_n}=0\quad \mbox{for}\ n=2k.\label{clofin}
\end{equation}

We see, that closeness provides further restrictions on the star product.

An interesting property of the gauge transformations with constant coefficients is that they
are, in fact, \emph{rigid} transformations. Therefore, they do not require introduction of any
new gauge fields to become symmetries of the action. Let us consider a classical action
\begin{equation}
S=\int d\mu (x) P(f_i,\nabla)_{\star_N}\,,\label{SN}
\end{equation}
where $f_i$ are some fields, $P$ is a polynomial, where all products are $\star_N$ products.
We can rewrite $S$ as
\begin{equation}
S=\int d\mu (x)D^{-1}( P(Df_i,\nabla)_{\star_R})=
\int d\mu (x)P(Df_i,\nabla)_{\star_R}\,.\label{SSR}
\end{equation}
This means, that the replacement $\star_N$ by $\star_R$ is compensated by the transformation
$f_i\to Df_i$. Since the operator $D$ is invertible, the theories based on the two star products
are classically equivalent.

\section{The metric}\label{sec-met}
Let us discuss how one can incorporate metric in the approach studied in this paper.
The most obvious choice to request the metric to be consistent with the same symplectic
connection, $\nabla_\mu g_{\nu\rho}=0$, but this imposes too sever restrictions on the
structure involved: because of the flatness of $\nabla$ it allows flat metrics only.
A general discussion of the compatibility conditions on the Riemann and Poisson
structures can be found in \cite{Hawkins}.

Let us reiterate the observation made in \cite{Vassilevich:2009cb}: the use of diffeomorphism
covariant star products allows to construct gravity theories which are invariant with
respect to the standard diffeomorphisms transformations without the need to make them twisted.
In particular, noncommutative counterparts of all 2D dilaton gravities \cite{Grumiller:2002nm}
were constructed in \cite{Vassilevich:2009cb} using the star product (\ref{dvstar}). These
models contain a complex zweibein $e_\mu$, a spin-connection, a dilaton, and auxiliary fields.
Apart from the diffeomorphisms, these models are invariant under noncommutative $U(1)_\star$
gauge transformations, which are deformations of the Euclidean Lorentz symmetries:
\begin{equation}
\delta e_\mu =i\lambda \star e_\mu ,\qquad
\delta \bar e_\mu =-i\bar e_\mu \star \lambda,\label{Lotr}
\end{equation}
The metric
\begin{equation}
g_{\mu\nu}=\frac 12 (\bar e_\mu \star e_\nu +
\bar e_\nu \star e_\mu),\label{gmn}
\end{equation}
is real and invariant under the $U(1)_\star$ transformations.

One of the models, corresponding to the conformally transformed string black hole,
appeared to be integrable \cite{Vassilevich:2009cb}.
The solution for the zweibein in that model reads:
\begin{equation}
e_\mu =u\star \nabla_\mu E,\qquad \bar e_\mu = \nabla_\mu \bar E \star u^{-1}\,,
\label{sole}
\end{equation}
where $u$ is a $U(1)_\star$ field which is canceled in the metric (\ref{gmn}) and can be
gauge-fixed to $u=1$. $E$ is an arbitrary complex scalar field corresponding to the
diffeomorphism freedom. In the coordinates $z^1={\rm Re}\, E$, $z^2={\rm Im}\, E$
the zweiben is a constant unit matrix. (In the commutative case this led to trivialization
of the geometry).
However, these coordinates $(z^1,z^2)$ need not
be the Darboux coordinates of the symplectic structure. Therefore, the geometric structure
on the noncommutative model may be very nontrivial.

Since the system possesses a full diffeomorphism invariance, the solutions may be analyzed
in any coordinate system. It is convenient to choose the  Darboux coordinates, where the
star product looks particularly simple (it coincides with the Moyal product):
\begin{equation}
\alpha \star \beta = \exp (i\theta (\partial_1^x \partial_2^y - \partial_2^x\partial_1^y)
\alpha (x)\beta(y)\vert_{y=x},. \label{Mstar}
\end{equation}
Here $\theta$ is a constant coefficient. We are using the
physical units with an imaginary deformation parameter ($h=i\theta$).

To give an example of possible behavior of the metric,
let us choose the arbitrary function $E$ in the form
\begin{equation}
E=\sin (x^1) + i \sin (x^2)\,. \label{Ess}
\end{equation}
This form is simple enough to allow explicit calculation of the star products.
On the other hand, it gives rise to a rich geometric structure.
The metric is easy to calculate
\begin{equation}
g_{\mu\nu}=\left( \begin{array}{cc}
\cos^2 x^1 & -\sin \theta\, \sin x^1 \sin x^2 \\
-\sin \theta\, \sin x^1 \sin x^2 & \cos^2 x^2 \end{array}\right) \,. \label{metE}
\end{equation}
The determinant of this metric
\begin{equation}
\det g_{\mu\nu}=\cos^2 x^1 \, \cos^2x^2 - \sin^2 \theta\, \sin^2 x^1 \, \sin^2 x^2
\label{detg}
\end{equation}
is positive for small values of $x^1$ and $x^2$, but changes the sign as $x^1$ and
$x^2$ grow!

To describe the geometry at small $x^1$ and $x^2$ it is convenient to introduce new
coordinates $z^1=\sin x^1$, $z^2=\sin x^2$. In these coordinates the line element
takes the form
\begin{equation}
({\mathrm{d}}s)^2=(\mathrm{d} z^1)^2+(\mathrm{d} z^2)^2 -\sin\theta\, f(z^1)f(z^2)
\mathrm{d} z^1\,\mathrm{d} z^2\,,\label{ds2}
\end{equation}
where $f(z)=z/\sqrt{1-z^2}$. To identify the geometry corresponding to (\ref{ds2})
one should calculate a diffeomorphism invariant corresponding to (\ref{ds2}).
The commutative Riemann tensor reads
\begin{equation}
R^1_{\ \ 212}=-\sin\theta\, f'(z^1)\, f'(z^2)\label{Riem}
\end{equation}
where we used the approximation of small $\theta$, which is valid for small $z^1,z^2$.
It is easy to see, that in this region, $f'\simeq 1$,
one has a constant curvature space. The sign of  the curvature is defined by the sign
of the noncommutativity parameter $\theta$.

We see, that the geometry obtained is indeed very non-trivial. The structure of the
solution is not described by the Cartan sector ($e_\mu$) alone (which is gauge trivial),
and not by the Poisson sector (the star product) alone (which is reducible to Moyal in
some coordinates), but rather by a relation between these two sectors.

A similar result has been obtained recently in \cite{Asakawa:2009yb}, where it was shown
that even a very simple action (just a cosmological constant) leads to a large variety of
the solutions on a noncommutative plane.

Therefore, to construct a noncommutative gravity theory it is not enough to present an
action for the gravity sector. One has to formulate an action principle which 
restricts\footnote{A relation between the metric and the symplectic structure can
be derived from the matrix models \cite{Steinacker:2008ya}.} also
the symplectic structure $\omega^{\mu\nu}_R$ and the symplectic connection $\nabla$. Fortunately,
since the Kontsevich gauge transformations with constant $l^{\mu_1\dots\mu_n}$ lead to classically
equivalent theories (see Sec.\ \ref{sec-clo}), there are no more relevant parameters in the
star product. Otherwise, one should have introduce a dynamical equation for each parameter.
This also implies, that a theory based on the generic star product with unrestricted gauge
freedom requires an infinite number of equations of motion, which makes such a theory meaningless.

A similar problem exists also in the approach to noncommutative gravities based on twisted
diffeomorphism symmetries \cite{tdiff}. In this approach, one has to pre-fix a relation
between the metric and noncommutativity, or between the star product and the Killing vectors
of the metric, see \cite{tsol}.
\section{Conclusions}\label{sec-con}
In this paper, we have analyzed covariant star products on the space of tensor fields over
a Fedosov manifold with a given symplectic structure and a given flat torsionless symplectic
connection. We have demonstrated, that although the space of star products has an infinite
functional dimension, this space can be reduced to a space of functional dimension one
after imposing some natural conditions (Theorem \ref{sec-gf}.1). The products $\star_N$
satisfying these conditions, and, therefore, called natural star products, are constructed
 in the following way. First, take the symplectic structure $\omega^{\mu\nu}$ and add
 arbitrary covariantly constant terms $\omega^{\mu\nu}_j$ multiplied with $h^{2j}$,
 see (\ref{Rom}). Then, with this new symplectic structure $\omega_R^{\mu\nu}$ one constructs
a star product $\star_R$, which is nothing else then a covariantization of the Moyal product.
Physical interpretation of this procedure of constant renormalization of the symplectic structure
is clear. Since we are working in the framework of formal expansions, there is no way
to ascribe a numerical value to the deformation parameter $h$. In a more physical setup, $h=i\theta$,
where $\theta$ is measurable, at least in principle.
One can sum up the expansion (\ref{Rom}) and impose on $\omega_R^{\mu\nu}$ a normalization
condition, as is usually done in quantum field theory. For example, this may be
$\omega_R^{\mu\nu}=\omega^{\mu\nu}$ for the physical value of the deformation parameter.
The second step consists in the Kontsevich gauge transformation (\ref{starN}) of $\star_R$
with covariantly constant coefficients in front of the differential operators contained in $D$,
which led us finally to $\star_N$.
The geometrical meaning of this transformation and its physical interpretation remain unclear.

We have studied some basic properties of the natural star products. We have shown that
all such products can be realized through a twist on a Hopf algebra and presented an explicit
construction of the twist element. We demonstrated that the closeness of the star product imposes
further restrictions on the parameters of the gauge transformations. Furthermore, we have
shown that classically field theories based on $\star_R$ and $\star_N$ are equivalent. This
means that classically $\omega_R$ and $\nabla$ are the only relevant parameters in the star product.

Let us now discuss the meaning of conditions S1-S6 (see Sec.\ \ref{sec-sta}) which we imposed on the star product. The conditions S1 -- S3 (existence of the formal expansion, associativity, and relation to the Poisson structure)  are rather straightforward extensions to all tensors of the
conditions which are always imposed on the star products of scalar fields in the deformation quantization approach. Later on we request that the Poisson bracket is common for all tensors,
which means that the noncommutative structure of the manifold is universal and does not depend
on the rank of the tensor -- an analog of  universality of the metric in General Relativity. The condition S5 (the Moyal symmetry) ensures hermiticity of the star product for physics (imaginary)
values of the deformation parameter $h$. The condition S4 (stability on covariantly constant tensors) means that slowly varying fields must not feel noncommutativity of the manifold. From the
mathematical point of view, S4 is a generalization of the condition of stability of unity. The condition
S6 (the derivation), which includes derivatives, relates the star products calculated in 
nearby points  (since $\nabla$ may be considered as a generator of  infinitesimal translations).
S4 and S6 are important properties of local products which seem to be natural to keep in
the nonlocal (noncommutative) case as well. These conditions appeared to be almost the same
restrictive as locality, as we demonstrated above.

Covariant star product were also studied in \cite{ACG,HoM,McCZ,Liev,tenv,Chaichian:2009kn}. The
aim of these works was to make the star products as general as possible. Therefore, the 
restrictions considered above were not imposed. An interesting manifestly covariant construction
of a star product of scalar functions was proposed in \cite{Cornalba:1998kt}. Finally, other types
of noncommutativity, as the ones following from matrix algebras, may also be covariantized and
extended at least to differential forms \cite{Madore}.

We have also considered a 2D gravity model \cite{Vassilevich:2009cb} based on the covariant star product.
Despite the gauge triviality of the zweibein, the metric in this model exhibits a very non-trivial
(and, in fact, rather wild) behavior. This simple example has demonstrated that the parameters in the
star product should be fixed by independent equations of motion -- and this is only possible if the
number of local parameters is reduced, as, for example, in the approach advocated in the present
work.

There is a number of possible further developments of the approach reported here. The most important
one seems to be an extension of the scheme to arbitrary Poisson manifolds.

\ack
I am grateful to Shannon McCurdy and Markku Oksanen for correspondence regarding covariant star products.
This work was supported in part by CNPq and FAPESP.

\appendix
\section{Commutative products}\label{appA}
Let us describe briefly commutative deformations of the point wise product. In this case all calculations are
considerably simpler. We put $\omega^{\mu\nu}=0$ (at all orders of $h$), so that the manifold $M$ is no longer a symplectic one.
Nevertheless, the product (\ref{dvstar}) exists and is just the usual point wise product $\alpha \cdot \beta$.
Let us now fix a flat torsionless connection $\nabla$, which is no longer restricted to being consistent with
any symplectic structure. We may define a new non-local covariant product as
\begin{equation}
\alpha * \beta = D^{-1} ((D\alpha)\cdot (D\beta))\,,\label{comstar}
\end{equation}
where $D$ is precisely as in (\ref{Dtau}). This product is obviously associative, so that S1 and S2 are satisfied,
and S3 defines a trivial Poisson structure. Next, we impose S4, which requires stability of the product on 
covariantly constant tensors. By repeating the calculations from the main text, we easily reproduce Eq.\ (\ref{res4})
which means that all $L^{\mu_1\dots\mu_n}$ are proportional to
unit operators on the spaces of tensors of a fixed rank. Now we
may conclude that the product (\ref{comstar}) is commutative. Although the inner automorphisms (\ref{inaut})
act trivially if $\star =\cdot$, the point wise product is diffeomorphism invariant, and we can exclude from the
gauge transformations all $L$ having a non-zero first order part $L^\mu$. The analysis of the higher order terms
proceeds exactly as above yielding the condition (\ref{Lncc}). 

The integration measure (\ref{intme}) is not defined in the commutative case, but it is enough to take 
$d\mu(x)=dx$ in a coordinate system, where $\nabla = \partial$. Then we may repeat the analysis of Sec.\
\ref{sec-clo} and conclude that all admissible products obtained by gauge transformations from the
$*$-product (\ref{comstar}) correspond to classically equivalent actions. 
In other words, if one relaxes the locality assumption but imposes instead the set of "natural" conditions
described above, the resulting field theories are classically equivalent to that with the local point wise
product.

\end{document}